\documentclass[12pt]{article} \usepackage{amssymb,amsfonts,latexsym}
\usepackage[dvips]{graphics} \usepackage{epsfig}
\def\complex       {{\dl C}}
\def\cft           {conformal field theory}
\def\Cft           {Conformal field theory}
\def\cfts          {conformal field theories}

\def\dl            {\mathbb }
\def\eE            {{\rm e}}
\newcommand\hsp[1] {\mbox{\hspace{#1 em}}}
\def\ii            {{\rm i}}
\def\infdim        {infinite-dimensional}
\newcommand\nxt[1] {\\\raisebox{.12em}{\rule{.35em}{.35em}}\hsp{.6}#1}
\def\reals         {{\dl R}}
\def\rep           {representation}
\def\Spec          {{\rm Spec}}
\def\twodim        {two-di\-men\-si\-o\-nal}
\def\zet           {{\dl Z}}

\begin{document}

\begin{flushright}  {~} \\[-1cm]
{\sf hep-th/0105266}\\{\sf PAR-LPTHE 01-26}
\\[1mm]
{\sf May 2001} \end{flushright}

\begin{center} \vskip 14mm
{\Large\bf THE WORLD SHEET REVISITED}\\[20mm]
{\large Christoph Schweigert$\;^1$ \ and \ J\"urgen Fuchs$\;^2$ 
}
\\[8mm]
$^1\;$ LPTHE, Universit\'e Paris VI~~~{}\\
4 place Jussieu\\ F\,--\,75\,252\, Paris\, Cedex 05 \\[5mm]
$^2\;$ Institutionen f\"or fysik~~~~{}\\
Universitetsgatan 1\\ S\,--\,651\,88\, Karlstad
\end{center}
\vskip 18mm
\begin{quote}{\bf Abstract}\\[1mm]
We investigate the mathematical structure of the world sheet in 
two-dimensional conformal field theories.
\end{quote}

\newpage

\section{On to the world sheet}

One way physicists think about conformal field theory is in terms of
a sigma model, defined as a ``quantization'' of a space of maps
from a \twodim\ world sheet $\Sigma$ to a target space $M$. Much effort has
been spent making the idea of quantization more precise.
The notion of vertex (operator) algebras is one result of these efforts. 
It captures basic features of a conformal field theory associated to 
$M$ (e.g.\ a free boson model when $M$ is flat) and thereby some aspects of 
the target space $M$ itself that are relevant for the quantized theory. In this
note we want to show that it is worthwhile and, for a deeper understanding of
conformal field theory, even indispensable to think about the 
structure of the world sheet $\Sigma$ as well.

A natural starting point for the discussion of $\Sigma$ is the 
the structure of a real, \twodim\ manifold. 
Such a manifold can have various additional structure:
\nxt it can be smooth
\nxt it can be compact
\nxt it can be orientable, and if so, it can be oriented
\nxt it can have boundaries
\nxt it can have a conformal structure
\nxt if it is orientable, it can have a holomorphic structure
\nxt it can have metric, with two possible choices of the signature
\nxt it can have a (generalized) spin structure

In different parts of the physics and mathematics literature on
conformal field theory different structures are needed. In this note
we present an attempt to clarify their relationship. From the outset,
the reader should be aware of one basic feature: There is nothing like 
``CFT$^\copyright$''. On the contrary, the term conformal field theory 
is used to refer to various different physical situations, 
and different purposes can and do require different axiomatizations.

Let us start with the question whether a metric, if present, has
Euclidean, i.e.\ $(++)$, or
Lorentzian, i.e.\ $(+-)$, signature. If we wish to think about conformal
field theory as a (toy) model for four-dimensional, ``realistic'' quantum
field theories, we are tempted to require a Lorentzian signature.
In fact, a lot of work has been done in this setting
(for a review see \cite{muge7}).

But -- in contrast to the situation in the Euclidean case --
the requirement of the existence of a metric of Lorentzian signature
severly restricts the topology of $\Sigma$. Indeed, 
a compact manifold admits a metric of Lorentzian signature if and only
if it admits a line element. This is the case precisely if its Euler 
characteristic vanishes. In two dimensions, 
a metric of Lorentzian signature is therefore an interesting
structure only when the manifold is either non-compact or when it is
a torus, which indeed can arise as a compactification\,% 
 \footnote{~This is not the conformal compactification, which is obtained by
 adding three points at future and past timelike as well as spacelike
 infinity, but a separate compactification of two light-like directions.}
of \twodim\ Minkowski space \cite{Mack}. We will not study this type of CFT 
in the remainder of this note.

Most `modern' applications of CFT indeed require a compact world sheet.
This is more or less evident in the application to \twodim\ critical
phenomena, since most samples have finite size. (It is worth noting
that compactness persists in the scaling limit that should be used to
investigate universal aspects of critical phenomena.) 

With regard to the application to string theory, compactness requires a 
few more words of explanation. The following naive picture is frequently 
suggested.  A string moves in a target space $M$ of Lorentzian signature. It
sweeps out a world sheet, which is endowed with the induced metric and thus 
has Lorentzian signature. 
The scattering of two ingoing particles to two outgoing particles
is then described (at the so-called tree level) by a picture of the
following form:
\vskip 4.7cm
$$
\begin{picture}(3,1)(100,0) %(width,height) position of origin
\scalebox{.3}{\includegraphics{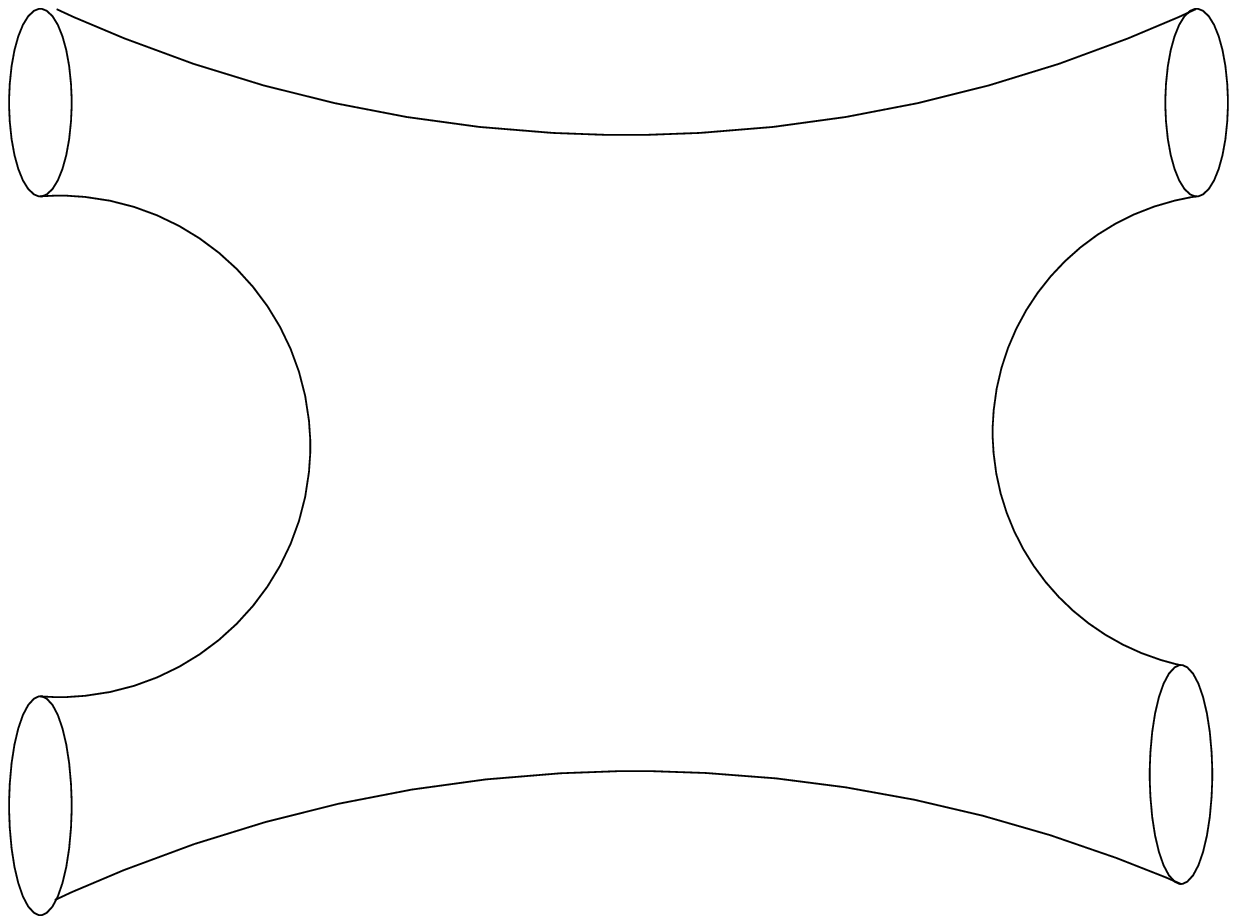}}
\end{picture}
$$
\mbox{$\ $}\\[-3.1cm]
At first sight this diagram suggests {\em non\/}-compactness of the
world sheet. But this first impression is misleading. One way to see this 
is to understand the real meaning of the ``boundaries'' of
the world sheet in this picture, which are to be thought of as parametrized 
circles. In fact, they are not physical boundaries, but rather their role is
to indicate 
the presence of ``asymptotic states''. Those, in turn, correspond to the 
insertion of suitable vertex operators on the world sheet. An appropriate way 
to interpret the situation is therefore to 
obtain local coordinates around each insertion of a vertex operator,
which can be achieved by gluing a standard disk
  $$ D = \{ z\,{\in}\,\complex \,|\, |z|\,{\leq}\, 1 \} $$
to each of the parametrized circles.

We will use such local coordinates instead of parametrized boundaries.
We then arrive at a compact world sheet. This leads to a conflict with 
a possible Lorentzian signature of a metric on the world sheet; we avoid 
it by always considering {\em Euclidean\/} metrics on the world sheet.
Thus we must give up the idea that the metric on the world sheet is 
related to the pull back of a metric on 
the target space. Note, however, that we do stick to a Lorentzian 
signature for target space\,% 
 \footnote{~This statement applies to the bosonic string and the
 $N\,{=}\,1$ superstring. The $N\,{=}\,2$ string requires a 
 four-di\-men\-si\-o\-nal target space of signature $(+,+,-,-)$.}.

This point of view is consistent with the following, somewhat minimalistic,
view of string theory: It constitutes a perturbation theory whose combinatorics
is captured in terms of surfaces and not, as in the case of usual Lagrangian
quantum field theory, in terms of graphs, the Feynman diagrams. A useful
analogy is supplied by the manner in which Kontsevich's prescription
for deformation quantization \cite{kont11} can be visualized in terms of 
amplitudes of a topological string theory on a disk \cite{cafe}.

It turns out to be crucial to distinguish the `fake' boundaries encountered 
above from real physical boundaries of world sheets. A first, mathematical, 
distinction is provided by the fact that the physical
boundaries are not parametrized. To
understand the difference between physical boundaries and fake boundaries
more clearly, it is instructive to consider some applications of 
physical boundaries:
\nxt The first application is provided by theories of open strings, which
     appeared already in the early days of string theory. They arose in an
     attempt to describe strong interactions, interpreting mesons, i.e.\
     particles composed of a quark and an antiquark, as open strings
     with the two charges that correspond to quarks and antiquarks
     attached to the end-points of the string.
     An open string is an interval; when it moves through space-time,
     it sweeps out a real two-dimensional surface $\Sigma$ with a boundary
     that is swept out by the end-points of the interval.
     Thus in this picture each boundary component
     corresponds to the world line of a charged particle.
\nxt In the application of conformal field theory to defects in solid state 
     physics, points on the boundary of the world sheet carry defects while
     the interior models the medium in which the defect is situated. 
     Thus, again the boundary is the world line of a
     particle-like defect with definite physical properties. \\
These examples indicate that physical boundaries have 
measurable physical properties 
and cannot be disposed of by gluing small disks to them. Moreover, these
properties can change along a component of a boundary. As a consequence,
physical boundaries can themselves support insertions, at which so-called 
boundary fields can change physical properties of the boundary.

We are thus led to study compact real surfaces, possibly with
boundaries, with a Euclidean metric
on it. Actually, a metric is a bit too much of a structure: We did not
take into account the conformal symmetry so far. The presence of this symmetry
implies that only the conformal class of the metric matters. That is, metrics 
$g$ and $g'$ that differ by a local rescaling,
$$ g'(p) = \eE^{\phi(p)} g(p) $$
with $\phi$ a sufficiently well behaved function on $\Sigma$, should be
identified. To be precise, we must be slightly more careful: If the Virasoro
central charge does not vanish, the so-called Weyl anomaly forces us to fix 
even a projective structure on $\Sigma$. (A projective structure on a
Riemann surface is an equivalence class of coverings by holomorphic
charts such that all transition functions are M\"obius transformations,
i.e.\ fractional linear transformations.)

Typically, many inequivalent conformal structures exist for a given 
topological manifold. They are parametrized by a moduli space. 
On can therefore consider ``one and the same'' conformal field theory
on a {\em family\/} of different spaces for the world sheet. This 
important idea does not have any direct analogue in conventional 
quantum field theory.

In string theory, we must go even further: Since no conformal structure
is preferred, the choice of a conformal structure must be regarded as an
auxiliary datum and hence be eliminated. Accordingly, scattering
amplitudes in string theory are defined as integrals of conformal field 
theory correlation functions over the moduli space of conformal 
structures. In this context, the following observation becomes relevant:
The moduli space is typically not compact, but
it can be compactified by including manifolds with singularities
that are not too bad. It is therefore common to consider conformal
field theories also on spaces that are more general than smooth manifolds and
that in particular can possess ordinary double points.  
{}From a purely mathematical point of view, the idea to
consider conformal field theories in families over moduli space and to
extend them to singular manifolds has been successful, too; the corresponding
factorization rules form the basis of most approaches to the Verlinde
formula \cite{beau,falt,tsuy}.

\section{Doubling the world sheet}

As we have seen, world sheets that are 
surfaces with boundaries should also be studied. Furthermore,
string theories of type I suggest that we should include unorientable
surfaces like the Klein bottle or the M\"obius strip in our discussion as 
well. CFT on the Klein bottle may seem somewhat exotic, and 
you might decide to resctrict your attention to orientable surfaces. But even
when you do so, this last remark should at least draw your attention to 
the fact that for the surfaces considered in the previous section
we did not choose an orientation.

Physicists often talk about left moving and right moving modes, which
suggests that both possible orientations should
be considered simultaneously. If the world sheet $\Sigma$ has empty
boundary (we will assume this in the next few paragraphs), this 
motivates us to consider along with $\Sigma$ also the total 
space $\hat\Sigma$ of its orientation bundle. Recall that the orientation
bundle is a $\zet_2$ principal bundle, where the two points in the fiber
correspond to the two possible choices of a local orientation. 
For example, the total space of the orientation bundle of a Klein bottle
is a torus $\complex/\zet+\ii t\zet$ with $t$ real that is a twofold covering 
of the Klein bottle; the Klein bottle is obtained by identifying the
points $z$ and $1\,{-}\,\bar z\,{+}\,\frac{\ii t}2$. When $\Sigma$ is
orientable, the total space of the orientation bundle consists of 
two copies of $\Sigma$ endowed with opposite orientation.

Let us describe the geometry of this space in more detail. $\hat\Sigma$
forms a twofold cover over $\Sigma$ \cite{ales,bcdcd}. 
This cover has two disjoint sheets
if and only if $\Sigma$ is orientable. Interchanging the two sheets
defines an anti-conformal involution $\sigma$ on $\hat\Sigma$, i.e.\
a map that reverses the orientation and preserves (the modulus of) 
angles. $\Sigma$ can be obtained from $\hat\Sigma$ as a quotient under
this action:
\begin{equation}
 \Sigma = \hat\Sigma / \sigma \, . \label{1} \end{equation}
In the physics literature this is referred to as a world sheet orbifold,
parameter space orbifold or (un-)orientifold \cite{clny2,poca,prsa}.
The total space $\hat\Sigma$ 
is not only orientable, it even possesses a canonical orientation.
On the other hand, it also inherits a conformal structure from $\Sigma$.
In two dimensions, together these data are equivalent to a complex 
structure on $\hat\Sigma$. 
This aspect of \twodim\ geometry supplies us with particularly powerful 
tools for the study of {\em two\/}-di\-men\-si\-o\-nal \cfts, as 
opposed to conformal field theories
in higher dimensions. In fact, we believe that this feature is far more
important than another peculiarity of two dimensions that is often
emphasized: the fact that the conformal algebra is \infdim.
 
The construction can be easily extended to surfaces with
boundary. We illustrate it in the case when $\Sigma$ is a disk: 
We only double the points in the interior of $\Sigma$, so that
$\hat\Sigma$ is a sphere obtained by gluing two disks to each other along 
their boundary. The statements about $\hat\Sigma$ made above then remain 
true, but now the anti-conformal involution does not act freely any longer;
its fixed points on $\hat X$ are in one-to-one correspondence with
the boundary points of $\Sigma$.

It is worth to pause at this point and to note that we have now arrived
at two different mathematical structures. First, a conformal real 
two-dimensional manifold $\Sigma$; second, its oriented cover $\hat\Sigma$, 
which is even a complex curve. (Both manifolds can in fact have singularities, 
a fact that, as already mentioned, we ignore for the moment.)

Correspondingly, there are actually two different types of theories that 
are referred to as conformal field theory: {\em Chiral\/} conformal field
theory, which lives naturally on a complex curve such as $\hat\Sigma$,
and {\em full\/} conformal field theory, which is defined on a real 
unoriented conformal manifold, possibly with boundary. 
The structure of a complex curve is frequently assumed in mathematical 
discussions of conformal field theory -- like in the definition of conformal
blocks, see e.g.\ \cite{BEfr} -- for the world sheet itself. Most of
the mathematical literature about conformal field theory therefore deals
with chiral conformal field theory. (Still, chiral CFT also has direct 
physical applications. In particular it describes universality classes of the 
edge system of quantum Hall fluids (for a review, see \cite{fpsw}).  

\section{Schemes and ringed spaces}

The relation between $\Sigma$ and $\hat\Sigma$ calls for a more conceptual 
explanation. As it turns out, it is convenient to regard the world sheet 
$\Sigma$ as a real 
scheme. The double is then just the complexification of this scheme:
\begin{equation}
\hat\Sigma = \Sigma \times_{\Spec(\reals)} \Spec(\complex) \,.
\label{2} \end{equation}

Let us consider a simple example to appreciate the situation -- the affine
scheme, given by the ring of polynomials in one variable over $\reals$
respectively $\complex$\,. Obviously, for the rings we have
\begin{equation}
\complex[X] = \reals[X] \otimes_{\reals} \complex \, . 
\end{equation}
The spectrum of (closed points 
of) $\complex[X]$ is well known: All non-trivial prime ideals 
are of the form $X\,{-}\,\alpha$ with $\alpha\,{\in}\,\complex\,$, 
hence $\Spec(\complex[X])$
is just the complex plane. The situation over $\reals$ is quite a bit
more subtle. Indeed, zeros of real polynomials are either real or come in
complex conjugate pairs. Thus the prime ideals are either of the form
$X\,{-}\,a$ with $a\,{\in}\,\reals$\,,
or $(X\,{-}\,\alpha)(X\,{-}\,\bar\alpha)$ with $\alpha\,{\in}\,
\complex\,{\setminus}\reals$\,. $\Spec(\reals[X])$ thus corresponds to
the complex plane with complex conjugate numbers identified, which is
topologically a half-plane.

Quite generally, a real scheme has complex points as well as real points. The
real points correspond to the boundary of $\Sigma$, the complex points
to the interior of $\Sigma$. 
Let us again consider the example of the ring of polynomials: A complex
point is a ring homomorphism to $\complex$\,, which for complex polynomials
amounts to the evaluation of the polynomial at some value of $X$.
For real polynomials, we can first of all consider real-valued points, i.e.\
ring homomorphisms to $\reals$\,. They correspond to the evaluation of
the polynomial at a real number. Evaluation at a complex number
gives a homomorphism to the $\reals$-algebra $\complex$\,, and complex
conjugate numbers give homomorphisms that are isomorphic over
$\reals$\,. The complex points in this example therefore correspond
to points in the interior of $\Sigma$, which are called ``bulk points''
in the physics literature. Finally, on the complexification we have
an action of the Galois group {\sl Gal}$(\complex,\reals)\,{\cong}\,\zet_2$,
whose non-trivial element is just the anti-conformal involution $\sigma$.

We can therefore summarize the situation concisely as follows. Chiral CFT 
lives on a complex curve, full CFT lives on a real curve. 
Accordingly, we wish to understand
how full CFT on a real curve is connected to some chiral CFT on its 
complexification. It turns out that this relation can be formulated in a 
model independent manner.
The mere fact that such a relation exists is good news indeed. It tells us
that all the mathematical
results about chiral conformal field theory, in particular the theory
of vertex operator algebras, are relevant for conformal field
theory on surfaces with boundaries as well.

Actually, this is not yet the end of the story. Certain classes of conformal
field theories require to endow
the world sheet with the structure of a scheme that is not a variety
and thereby explore the full power of schemes.\,%
 \footnote{~The idea to regard the world sheet as a ringed space, i.e.\ as
 a topological space with a sheaf of rings on it, has already appeared
 in a different context: In \cite{rain11}, ringed spaces with nilpotents
 have been used to describe normal ordering in specific classes of chiral
 CFTs, so-called $b$-$c$ systems.}
The best known example of such chiral conformal theories are superconformal 
theories. These are models of conformal field theory whose chiral algebra is 
a super vertex algebra \cite{KAc4}
(with a super-Virasoro vector, but this vector does not enter our discussion). 
In the physics literature it is standard lore that the introduction of 
fermionic fields in the chiral algebra requires the choice of a spin structure 
on the world sheet. String theory correlators are then constructed by
summing over all spin structures.

There are several ways of looking at a spin structure. The naive idea of 
choosing a square root $\mathcal S$ of the canonical bundle $\mathcal K$, i.e.\
$(\mathcal S)^{\otimes 2}\,{=}\,{\mathcal K}$, is not necessarily the best 
way. Namely, in the theory of simple currents, one would like to generalize 
the situation to orders $N$ higher than two. However, on curves of 
general genus $g$ the equation $(\mathcal S)^{\otimes N}\,{=}\,{\mathcal K}$
does not possess any solution, due to the integrality of the degree of a line
bundle. It is therefore better to recall that the category of spin 
curves is equivalent to the category
of supersymmetric curves. A supersymmetric curve, on the other hand, should
be seen as a ringed space with a sheaf of graded-commutative rings that
contain nilpotent elements. Such ringed spaces, in turn, allow for 
generalizations.

A second class of models 
in which the structure of the world sheet must be extended 
are orbifold theories based on some orbifold group $G$. It has been 
advocated already long ago \cite{hava} that conformal blocks in such CFT 
models can be computed using covering surfaces. Instead of $\hat\Sigma$, one
works with a (possibly branched) covering surface $\tilde\Sigma$ such that
  $$ \tilde\Sigma / G  = \hat\Sigma \, . $$ 
The similarity with (\ref1) is striking, and again we would like to
take it into account by extending the structure sheaf, as in (\ref2).
It therefore seems that the natural structure 
for the world sheet is the one of
a ringed space, where the structure of the ring has to be adapted to
the vertex algebra that underlies the theory. 

This structure may sound complicated. But fortunately the following
observation simplifies life: 
As shown in \cite{BAki} in the case of complex curves, the Riemann-Hilbert
correspondence implies that a complex modular functor can be described by an 
equivalent topological modular functor.
For many purposes, in particular for the construction 
of a full CFT from a chiral CFT, one can therefore work entirely
in a topological setting \cite{fffs3}. We expect 
that this pattern generalizes in such a way that the conformal 
field theory of orbifolds should find a natural topological counterpart in
so-called $\pi$-manifolds \cite{tura7}. 
This story, however, has yet to be unraveled.

\section{Chiral and full CFT}

Based on purely geometric considerations, we have been led to two
distinct types of conformal field theories:
\nxt Chiral conformal field theory, defined on closed 
     oriented surfaces.
\nxt Full conformal field theory, defined on unoriented (and possibly
     unorientable) surfaces which can have a boundary.

We conclude with a few comments on both classes of theories.
The protagonists of chiral conformal field theory are the conformal blocks.
In the algebro-geometric setting a conformal block corresponds to a certain 
vector bundle over the moduli space of curves; there is also a topological 
description as a vector space associated to an ``extended surface'' \cite{TUra}.
It is known \cite{BEfr} that to every vertex algebra one can associate
a system of conformal blocks. 

The following are, in our opinion, among the most pressing
mathematical questions about chiral conformal field theory:
\nxt Under which conditions on the vertex algebra do the conformal blocks
     possess good factorization properties?
\nxt Under which conditions is the tensor category of \rep s of a rational 
     vertex algebra modular?
\nxt What are the good notions that allow to understand non-rational theories,
     including so-called non-compact conformal field theories? \\[.1em]
(Note that one of the big virtues of the vertex algebras associated to certain
infinite-dimensional Lie algebras is 
their ability to select a ``good'' subcategory of the representations
of the Lie algebra, much like Lie groups do. 
This feature could become even more important in the study of non-compact 
conformal field theories that are based on non-compact forms of Lie algebras.
The representation theory of those Lie algebras is a rich subject in itself,
and it will be interesting to see what subcategories
of representations are chosen by the associated vertex algebras.) 

\smallskip

The protagonists of full conformal field theory are the correlation functions.
In the applications of conformal field theory to statistical mechanics
they encode physical quantities, such as scaling dimensions or critical
exponents, and in string perturbation theory their integrals
over moduli space provide scattering amplitudes. 

One of the important insights in \cft\ is that full CFT on a conformal 
surface $\Sigma$ is closely connected to chiral CFT on the complex curve 
$\hat\Sigma$. The situation is in fact as beautiful as one could have
hoped for: One can construct all correlation functions in terms of conformal 
blocks in an entirely model-independent
manner (see e.g.\ \cite{fffs3,fuSc14}). There is, however, still more to be
uncovered; some of the category-theoretic tools that are needed for
a complete understanding are
developed in the contribution \cite{fuSc15} to this volume. 

Owing to these constructions, chiral conformal field theory in general, and 
notably vertex algebras and their representation theory,
are in particular of direct relevance to the CFT
description of certain solitonic sectors of string theory, so-called
D-branes. These play an important role for various recent developments
in string theory such as duality symmetries, dynamics of supersymmetric
gauge theories, and black hole entropy.
It can therefore be expected that vertex algebras and related 
mathematical structures will continue to play an important role in 
string theory.

\vskip1.5em

\noindent
{\bf Acknowledgements} \\
We would like to thank W.\ Soergel for drawing our attention to the 
complexification of real schemes.

 \newcommand\wb{\,\linebreak[0]} \def\wB {$\,$\wb}
 \newcommand\Bi[1]    {\bibitem{#1}}
 \newcommand\J[5]   {{\em #5}, {#1} {\bf #2} ({#3}), {#4} }
 \newcommand\JL[4]    {{\bf #2} ({#3}), {#4} }
 \newcommand\Prep[2]  {{\em #2}, pre\-print {#1}}
 \newcommand\PRep[2]  {{\em #2}, {#1}}
 \newcommand\BOOK[4]  {{\em #1\/} ({#2}, {#3} {#4})}
 \newcommand\inBO[7]  {{\em #7}, in:\ {\em #1}, {#2}\ ({#3}, {#4} {#5}), p.\ {#6
}}
 \def\jf    {J.\ Fuchs}
 \def\adma  {Adv.\wb Math.}
 \def\aspm  {Adv.\wb Stu\-dies\wB in\wB Pure\wB Math.}
 \def\atmp  {Adv.\wb Theor.\wb Math.\wb Phys.}
 \def\coia  {Com\-mun.\wB in\wB Algebra}
 \def\coma  {Con\-temp.\wb Math.}
 \def\comp  {Com\-mun.\wb Math.\wb Phys.}
 \def\cpma  {Com\-pos.\wb Math.}
 \def\duke  {Duke\wB Math.\wb J.}
 \def\foph  {Fortschritte\wB d.\wb Phys.}
 \def\imrn  {Int.\wb Math.\wb Res.\wb Notices}
 \def\inma  {Invent.\wb math.}
 \def\injm  {Int.\wb J.\wb Math.}
 \def\jgap  {J.\wb Geom.\wB and\wB Phys.}
 \def\joag  {J.\wB Al\-ge\-bra\-ic\wB Geom.}
 \def\joal  {J.\wB Al\-ge\-bra}
 \def\jomp  {J.\wb Math.\wb Phys.}
 \def\jopa  {J.\wb Phys.\ A}
 \def\josp  {J.\wb Stat.\wb Phys.}
 \def\maan  {Math.\wb Annal.}
 \def\nuci  {Nuovo\wB Cim.}
 \def\nupb  {Nucl.\wb Phys.\ B}
 \def\phlb  {Phys.\wb Lett.\ B}
 \def\phrl  {Phys.\wb Rev.\wb Lett.}
 \def\slnm  {Sprin\-ger\wB Lecture\wB Notes\wB in\wB Mathematics}
 \def\rvmp  {Rev.\wb Math.\wb Phys.}
 \def\tams  {Trans.\wb Amer.\wb Math.\wb Soc.}
 \def\topo  {Topology}
   \def\AMS    {{American Mathematical Society}}
   \def\BIR    {{Birk\-h\"au\-ser}}
   \def\OUP    {{Oxford University Press}}
   \def\PL     {{Plenum Press}}
   \def\SV     {{Sprin\-ger Ver\-lag}}
   \def\Bo     {{Boston}}
   \def\PR     {{Providence}}
   \def\NY     {{New York}}

\end{document}